\documentclass{elsart}

\usepackage{amsmath}                   
\usepackage{amssymb}                   
\usepackage{graphicx}                  
\usepackage{subfigure}
\usepackage{rotate}
\begin{document}


\begin{frontmatter}
\journal{Astroparticle Physics}

\title{
Measurement of the Aerosol Phase Function at the 
Pierre Auger Observatory
}

\author[Columbia]{S.Y.~BenZvi},
\author[Columbia]{B.M.~Connolly},
\author[UNM]{J.A.J.~Matthews},
\author[Columbia]{M.~Prouza},
\author[Columbia,Carnegie]{E.F.~Visbal}, and 
\author[Columbia]{S.~Westerhoff}

\address[Columbia]{Columbia University, Department of Physics and Nevis Laboratories,
                   538 West $\it 120^{th}$ Street, New York, NY 10027, USA}
\address[UNM]{University of New Mexico, Department of Physics and Astronomy, Albuquerque, 
              NM 87131, USA}
\address[Carnegie]{Carnegie Mellon University, Department of Physics, Pittsburgh, 
                   PA 15213, USA}


\begin{abstract}
Air fluorescence detectors measure the energy of ultra-high energy cosmic rays by collecting
fluorescence light emitted from nitrogen molecules along the extensive air shower cascade.
To ensure a reliable energy determination, the light signal needs to be corrected for
atmospheric effects, which not only attenuate the signal, but also produce a non-negligible 
background component due to scattered Cherenkov light and multiple-scattered light.  The 
correction requires regular measurements of the aerosol attenuation length and the aerosol
phase function, defined as the probability of light scattered in a given direction.  At 
the Pierre Auger Observatory in Malarg\"ue, Argentina, the phase function is measured on an 
hourly basis using two Aerosol Phase Function (APF) light sources.  These sources direct a UV 
light beam across the field of view of the fluorescence detectors; the phase function can be 
extracted from the image of the shots in the fluorescence detector cameras.  This paper 
describes the design, current status, standard operation procedure, and performance of the 
APF system at the Pierre Auger Observatory.
\end{abstract}


\end{frontmatter}

\section{Introduction}\label{sec:intro}

The Pierre Auger Observatory in Malarg\"ue, Argentina, is designed to study the
origin of ultra-high energy cosmic rays with energies above $10^{18}$\,eV.  While 
still under construction, scientific data taking began in 2004, and first results
have been published~\cite{Mantsch:2005,Abraham:2006,Aglietta:2006}.

The Pierre Auger Observatory is a hybrid detector that combines two techniques 
traditionally used to measure cosmic ray air showers: surface particle 
detection and air fluorescence detection.  Both detector types
measure the cosmic ray primary indirectly, using the Earth's atmosphere as part of
the detector medium.  When the primary particle enters the atmosphere, it interacts
with air molecules, initiating a cascade of secondary particles, the so-called
extensive air shower.  Surface detectors in the form of ground arrays sample the 
shower front as it impacts the ground, whereas air fluorescence detectors make use 
of the fact that the particles in the air shower excite nitrogen molecules in the air, 
causing UV fluorescence.  Using photomultiplier cameras to record air shower UV
emission, we can observe showers as they develop through the atmosphere and obtain 
a nearly calorimetric estimate of the shower energy.

Upon completion, the surface detector (SD) array of the Pierre Auger Observatory 
will comprise 1600 water Cherenkov detector tanks, deployed in a hexagonal grid 
over an area of $3000~\mathrm{km}^{2}$, and four fluorescence detector (FD) stations 
overlooking the SD from the periphery.  An advantage of combining both detector 
types at the same site is the possibility to cross-calibrate.  Based on the subset 
of events seen with both detectors, the nearly calorimetric information of the FD 
provides the energy calibration of the SD.  

For the calibration to be meaningful, the properties of the calorimeter, {\it i.e.} 
the atmosphere, must be well-known.  At the Pierre Auger Observatory, this is achieved 
by an extensive program to monitor the atmosphere within the overall FD aperture 
and measure atmospheric attenuation and scattering properties in the 300 to 400~nm 
wavelength band recorded by the FDs~\cite{Cester:2005,BenZvi:2007,Fick:2006}.  

Two primary forms of atmospheric light scattering need to be considered: molecular, 
or Rayleigh, scattering, mainly due to nitrogen and oxygen molecules; and aerosol 
scattering due to airborne particulates.  The 
angular distribution of scattered light in both types of scattering may be 
described by a phase function $P(\theta)$, defined as the probability per unit 
solid angle of scattering through an angle $\theta$.

Rayleigh scattering allows for an analytical treatment, and assuming isotropic
scattering, the Rayleigh phase function has the well known $1+\cos^{2}\theta$ 
angular dependence.  Matters are more complicated for
aerosols, because the scattering cross section depends on the size
distribution and shape of the scatterers.  Forward scattering typically
dominates in this case, but the fraction of forward-scattered light
varies strongly with aerosol type.  Moreover, a rigorous analytical
treatment is not possible, though the literature gives various
approximations.  For example, if one assumes spherical particles with a
known or estimated size distribution, then aerosol scattering can be
described analytically using Mie theory~\cite{Mie:1908}.  In practice, however,
aerosols vary a great deal in size and shape, and the aerosol content of
the atmosphere changes on short time scales as wind lifts up dust,
weather fronts pass through, or rain removes dust from the atmosphere.

The FD reconstruction of the primary cosmic ray particle energy must account 
not only for light that is ``lost'' between the shower and the camera due to 
scattering, but also for direct and indirect Cherenkov light contributing to 
the FD signal.  The amount of Cherenkov light seen by the FDs depends on the 
viewing angle, {\it i.e.} the angle between the shower axis and the FD line 
of sight, and can be calculated once the geometry of the air shower is determined.  
At small viewing angles, direct Cherenkov light dominates, while at viewing angles 
greater than $\sim 20^{\circ}$, the FDs detect mainly ``indirect'' Cherenkov 
light scattered into the FD field of view.  To calculate this scattered component, 
the aerosol phase function needs to be known.  Finally, a small multiple scattering 
component also adds to the contamination of the fluorescence light and must be 
removed~\cite{Roberts:2005}.

The Aerosol Phase Function (APF) light sources~\cite{Matthews:2001,Matthews:2003}, 
in conjunction with the fluorescence detectors at the Pierre Auger Observatory, are 
designed to measure the aerosol phase function on an hourly basis during FD data taking.  
The APF light sources direct a near-horizontal pulsed light beam across the field of 
view of a nearby FD.  The aerosol phase function can then be reconstructed from the 
intensity of the light observed by the FD cameras as a function of scattering angle.
Since the FD telescopes cover about $180^{\circ}$ in azimuth, the aerosol phase 
function is measured over a wide range of scattering angles.  

Currently, APF light sources are installed and operating at two of the FDs.
With their ability to measure the {\it angular 
distribution} of the scattered light, the APF light sources are meant to complement 
other atmospheric monitoring tools at the Auger site which measure the optical 
depth, and therefore the {\it amount} of attenuation due to aerosols.

This paper describes the design and performance of the APF light sources.
It is structured as follows.  Section 2 gives a description
of the APF facilities.  Section 3 describes how the aerosol phase function 
is determined from the APF data.  In Section 4, we show first results for
data taken between June and December 2006.  Section 5 summarizes the paper.
 
\begin{figure}[t]
\centering
\includegraphics[scale=.5]{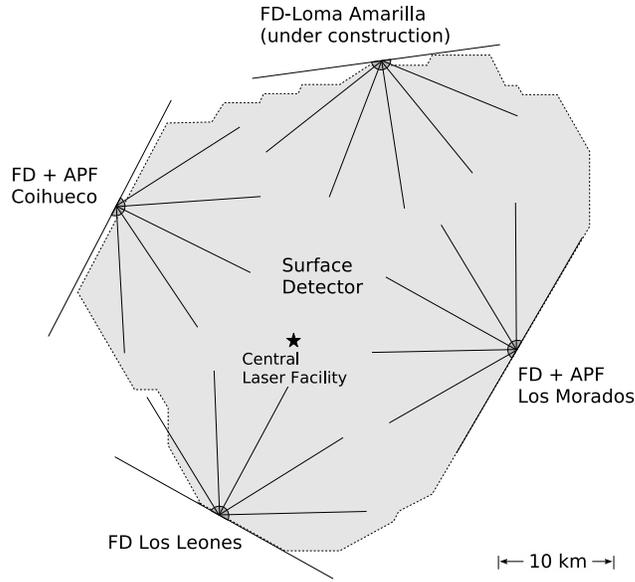}
\caption{\it Schematic layout of the Pierre Auger Observatory.  The shaded area 
indicates the shape and size of the surface detector area.  The fluorescence detectors
are placed at the periphery of the surface detector array.  The field of view of
the 6 bays of each fluorescence detector (FD) is indicated by the lines.
From the Central Laser Facility (CLF)~\cite{Fick:2006} in the center of the surface detector
array, a pulsed UV laser beam is directed into the sky, providing another test
beam which can be observed by the FDs.}
\label{auger_scheme}
\vskip1cm
\end{figure}

\section{APF Light Sources}

\subsection{Detector Buildings, Optics, and Electronics}

\begin{figure}[t]
\centering
\includegraphics[scale=.3]{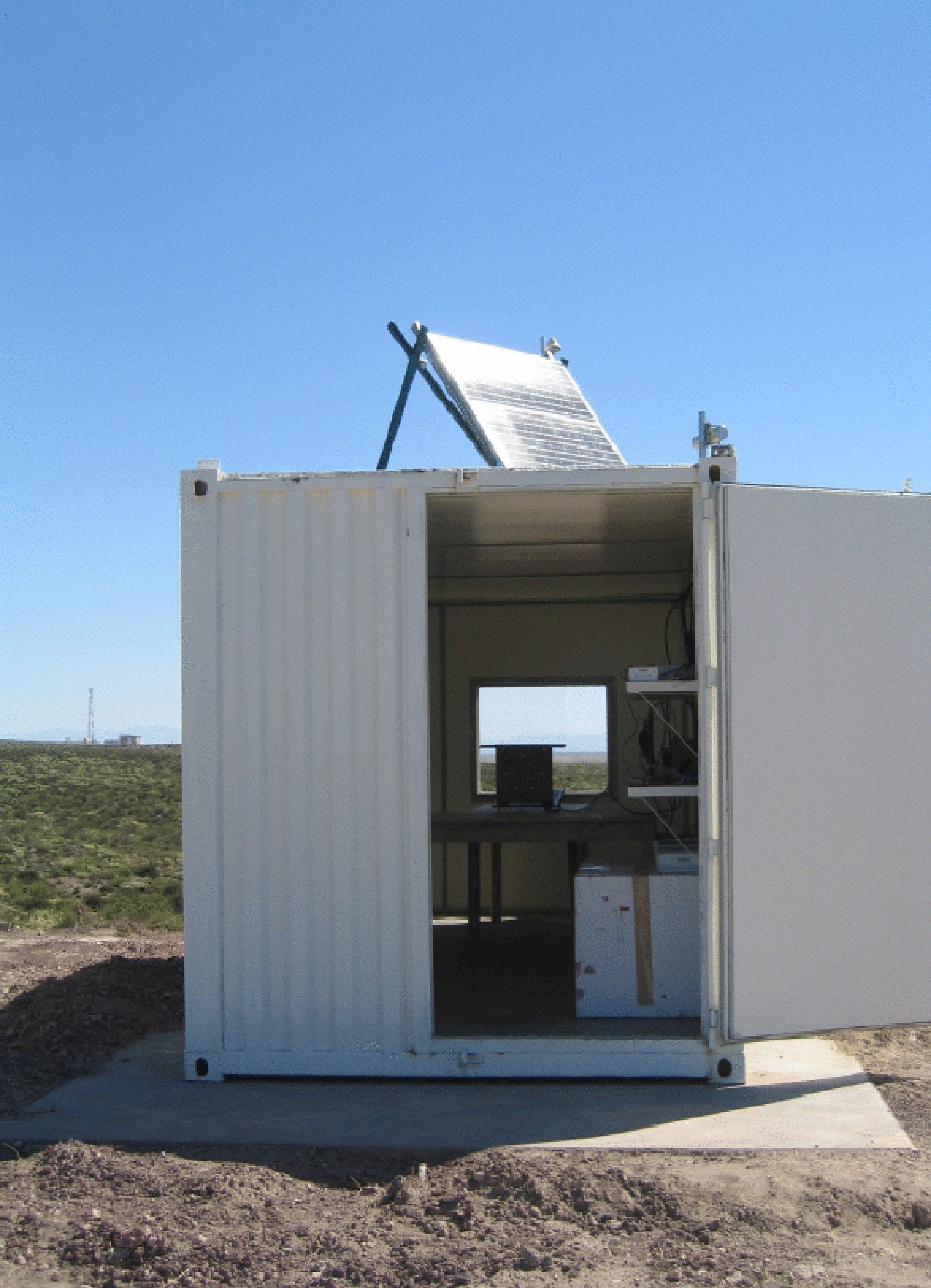}
\includegraphics[scale=.3]{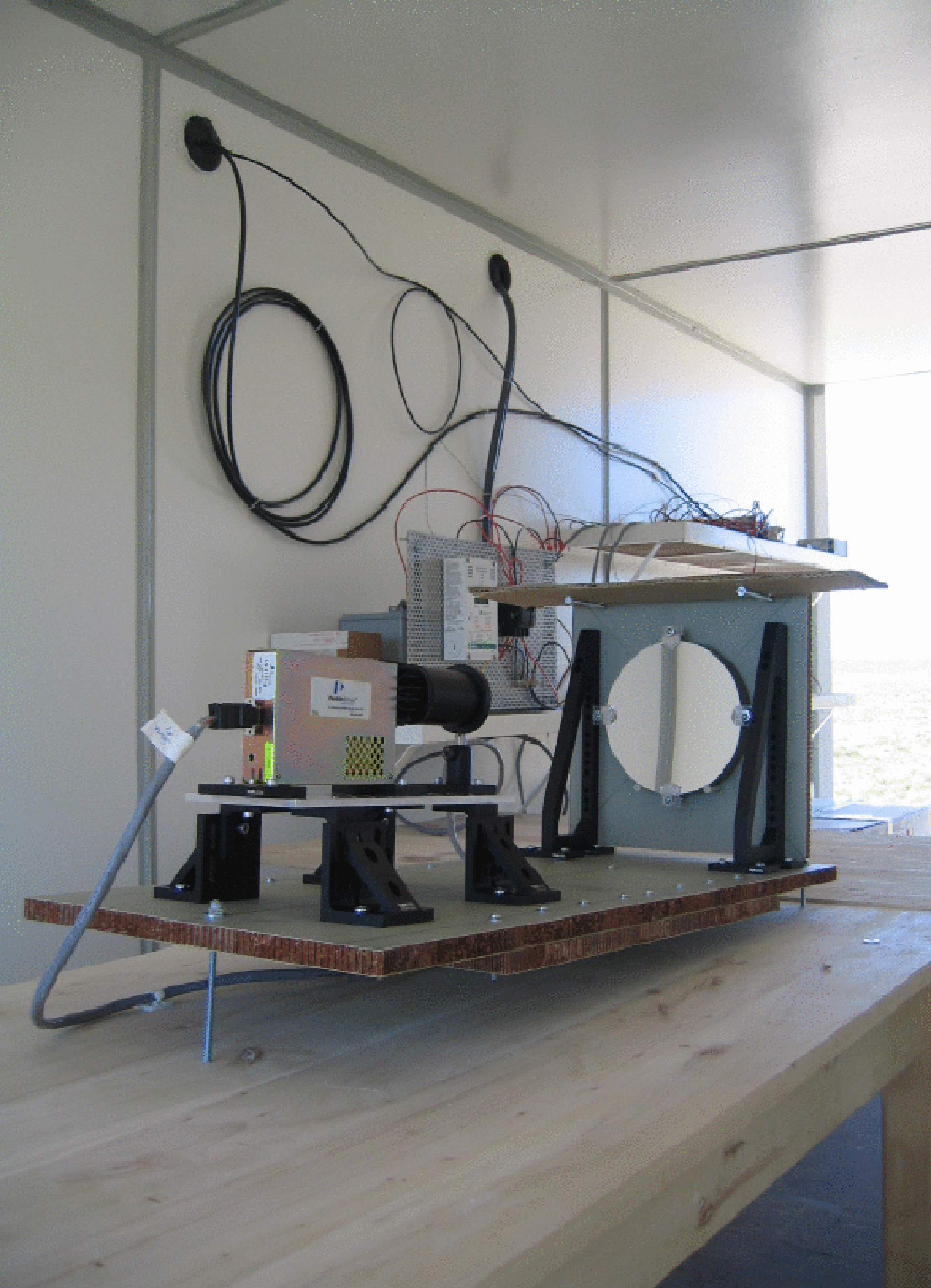}
\caption{\it Photos of the enclosure (left) and the light source (right) at the
Los Morados APF facility.  In the photograph on the left, the Los Morados FD can be seen 
on the horizon (to the left of the container).}
\label{apf_photo}
\vskip1cm
\end{figure}

The Auger FD comprises four detector stations (see Fig.\,\ref{auger_scheme}).
At present, the sites at Los Leones, Coihueco, and Los Morados are completed 
and fully operational, while the fourth site at Loma Amarilla is under 
construction.  APF light sources are operating at the Coihueco and Los Morados
FD sites.  Both were built by the University of New Mexico 
group~\cite{Matthews:2003}.  Fig.\,\ref{apf_photo} shows a photograph of the 
APF container building at Los Morados.

Each APF building contains sources which operate at different wavelengths in the 
region of interest between 300~nm and 400~nm.  During the initial studies described 
in this paper, only one light source with a Johnson U-band filter of central
wavelength 350~nm was used.  However, in the near future, we plan to 
operate the light sources at several wavelengths to study the wavelength dependence 
of the phase function over the full range of the FD sensitivity.

The light beam is provided by a broad-band Xenon flash lamp source from Perkin Elmer
Optoelectronics (model LS-1130-4 FlashPac with FX-1160 flash lamp).  The Xenon flash 
lamps were chosen because of their excellent stability in intensity and pulse shape.
A Johnson/Cousins (Bessel) U-band filter from Omega Optical Inc. (part number XBSSL/U/50R)
selects a central wavelength of $\sim 350$~nm, FWHM 60~nm) from the broad flash 
lamp spectrum.  The beam is focused using a 20.3~cm
diameter UV enhanced aluminum spherical mirror (speed f/3) 
from Edmund Scientific Co. (part number R43-589).
All optical components are assembled on a commercial optical plate.
We use Thor optical table parts, assembled from Nomex 
Epoxy/Fiberglass 1.91~cm panels from TEKLAM (part number N507EC).

The Xenon lamps rest inside refurbished 6.1~m
shipping containers, and the light is sent through a 0.749~cm
thick acrylite UV transmitting window (Cyro Industries acrylite OP-4 UVT acrylic).  
Each light source provides a nearly horizontal beam of divergence $\leq 10~\mathrm{mrad}$ 
pulsed across the field of view of the nearby fluorescence detector.
Computer control occurs from the corresponding FD building.  A serial radio link 
(YDI Wireless, model 651-900001-001 (TranzPoint ESC-II Kit)) connects the computer 
to a commercial ADC/relay system (model ADC-16F 16 channel 8 bit ADC and RH-8L 8-relay 
card from Electronic Energy Control Inc.) at the light source.  

Once during each hour of FD data taking, the ADC/relay system enables a 1~Hz GPS pulser 
(CNS Systems Inc., model CNSC01 with TAC32 software) and a 12~V to 24~V inverter 
to power the Xenon flash lamps.  Each lamp fires a set of 5 shots, pulsed at 2 second 
intervals.  The APF events are flagged by the FD data acquisition system and the 
corresponding FD data are stored on disk in especially designated APF data files.

When the light sources are not operating, only the radio link and the ADC board are powered.
The total current draw is therefore only $\sim 0.2\,\,\mathrm{A}$ at 12~V, and the whole 
system can be powered by batteries recharged during the day with 12~V solar panels 
(two Siemens SP75 75~W solar modules with Trace C35 controller).

\subsection{APF Signals in the Fluorescence Detectors}

The light beam produced by the APF sources is observed by the cameras of the corresponding FD
site.  The FD detectors of the Pierre Auger Observatory are described in detail 
elsewhere~\cite{Bellido:2005}.  Here, we only give a short summary of the main 
characteristics relevant for the analysis of APF shots.

Each Auger FD site contains six bays, and each bay encloses a UV telescope
composed of a spherical light-collecting mirror, a photomultiplier camera at the focal
surface, and a UV transmitting filter in  the aperture.  The mirrors have a radius
of curvature of 3.4~m and an area of about $3.5\times 3.5~\mathrm{m}^{2}$.  The camera
consists of 440 photomultipliers with a hexagonal bialkaline photocathode, arranged in a
$20\times 22$ array.  Each camera has a field of view of $30.0^{\circ}$ in azimuth and
$28.6^{\circ}$ in elevation, covering an elevation angle range from $1.6^{\circ}$ to
$30.2^{\circ}$ above horizon.  To reduce optical aberrations, including coma, the FD 
telescopes use Schmidt optics with a circular diaphragm of diameter 2.2~m placed at 
the center of curvature of the mirror, and a refractive corrector ring at the telescope 
aperture.

\begin{figure}[t]
\centering
\includegraphics[scale=.7]{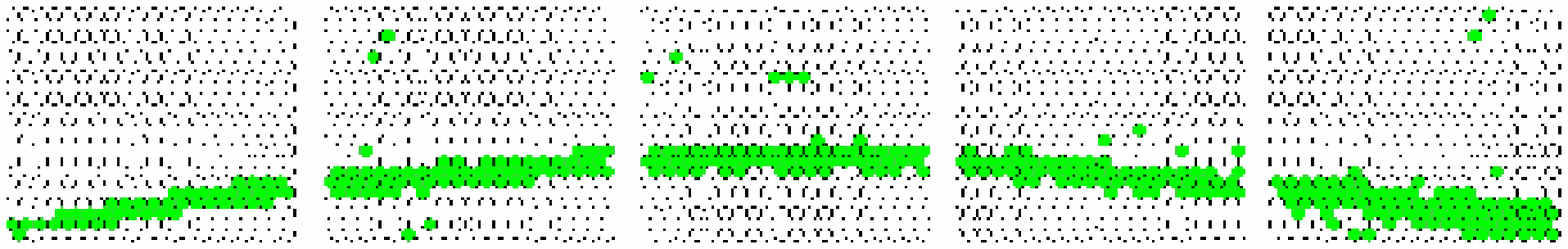}
\caption{\it The APF pulse as seen by the Coihueco FD.  The light travels from right to left,
and each PMT Cluster observes $30^{\circ}$ in azimuth.  Note that the projection of the 
approximately horizontal APF beam onto the spherical FD surface results in a curved track.}
\label{apf_shot}
\includegraphics[scale=1.]{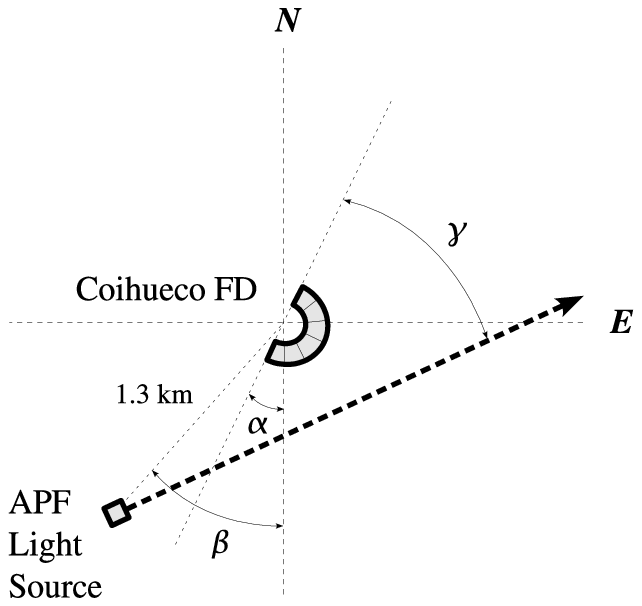}
\caption{\it Scheme of the location of the Coihueco APF light source relative to the 
Coihueco FD.  Located at the center is the Coihueco FD with its field of view indicated.
The value of $\alpha$ is $26^{\circ}$ and $\beta$ is $38^{\circ}$, measured from the North.
The shot direction $\gamma$ is about $24^{\circ}$.}
\label{apf_geo}
\vskip1cm
\end{figure}

Fig.\,\ref{apf_shot} shows an APF shot as seen by the Coihueco FD.  Five out of the 6 bays
of the Coihueco FD site observe light from the Coihueco APF facility.  In this figure,
the light travels from right to left.
Fig.\,\ref{apf_geo} shows the relative positions of the APF source and 
the FD at the Coihueco site.  The geometry is in part dictated by the local topography, and 
consequently is slightly different for the Los Morados site.

\section{Determination of the Aerosol Phase Function}\label{sec:method}

The signal from the APF light source observed by the $i^{th}$ pixel of a fluorescence detector 
can be expressed as

\begin{equation}
S_i = I_0 \cdot T_i \cdot \left[ \frac{1}{\Lambda_m} 
\left(\frac{1}{\sigma_m} \frac{d \sigma_m}{d\Omega}\right)+\frac{1}{\Lambda_a}
\left(\frac{1}{\sigma_a} \frac{d\sigma_a}{d\Omega}\right)\right] _i \cdot 
\Delta z_i \cdot \Delta \Omega_i \cdot \epsilon_i~~.
\label{eq:signal}
\end{equation}

In this equation, $I_0$ is the light source intensity; 
$T_i$ is the transmission factor $e^{-r_i/\Lambda_{tot}}$ which 
accounts for light attenuation from the beam to the pixel; 
$r_i$ is the distance from the beam to the detector; 
$\Lambda_{tot}$, $\Lambda_m$, and $\Lambda_a$ are the total, molecular, 
and aerosol extinction length, respectively; 
and $\sigma_m^{-1} d\sigma_m/d\Omega$ and $\sigma_a^{-1} d\sigma_a/d\Omega$ are the 
normalized differential molecular and aerosol scattering cross sections, respectively,
which are identical to the phase functions $P_m(\theta)$ and $P_a(\theta)$.  The integral 
of $P_m(\theta)$ and $P_{a}(\theta)$ over all solid angles is equal to 1.  Finally, 
$\Delta z_i$, $\Delta\Omega_i$, and $\epsilon_i$ are the track length, detector solid 
angle, and the efficiency for the $i^{th}$ pixel of the detector.  

The data come in the form of total PMT signal per pixel from a particular shot.  Those 
data are binned as a function of azimuth and averaged between the five shots taken within 
10 seconds.  In this analysis, $5^{\circ}$ bins are used, although the fit is relatively
insensitive to the number of bins.  Each FD pixel is hexagonally shaped, so for those
lying at the boundary of two azimuth bins, the fractional area of the hexagon in each bin 
is used to properly distribute the signal. The signal in each pixel is divided by 
$\Delta z_i$, $1/r_i^2$ and $\epsilon_i$ to correct for the geometry of the beam 
and pixel calibration.  Note that in the roughly cylindrical geometry of the FD-APF beam,
the $\Delta z_i$ and $1/r_i^2$ corrections almost completely cancel out.

Typical values for the aerosol extinction length in dry atmospheres are between
10~km and 20~km, reaching 40~km for very clear conditions.  Since the perpendicular 
distance from the beam to the FD is only on the order of a few hundred meters, it 
is reasonable to assume full atmospheric transmission $(T_i=1)$ over the length of 
the beam.  In reality, this assumption does not hold well for the most distant beam 
points, so these points are not used in the present study.  In the near future, 
measurements of the extinction length from the Auger lidar stations~\cite{BenZvi:2007}
will be used to improve the APF analysis.  In another approximation, we assume that 
the extinction lengths are identical for each pixel for single measurements and do 
not require an index $i$. In principle, the extinction length depends on the number 
density of scatterers and is therefore a function of the density (temperature, 
pressure) of the air. 

Given corrections for geometry, attenuation, and pixel efficiency,  Eq.\,\ref{eq:signal}
reduces to 
\begin{equation}
S_i = C \cdot\left[ \frac{1}{\Lambda_m} \left(\frac{1}{\sigma_m}
\frac{d \sigma_m}{d\Omega}\right)+\frac{1}{\Lambda_a}
\left(\frac{1}{\sigma_a}\frac{d\sigma_a}{d\Omega}\right)\right]~~,
\end{equation}
where $C$ is a constant whose value is unimportant because arbitrary units are sufficient 
in determining the phase function.  

\begin{figure}[t]
\centering
\includegraphics[scale=1.]{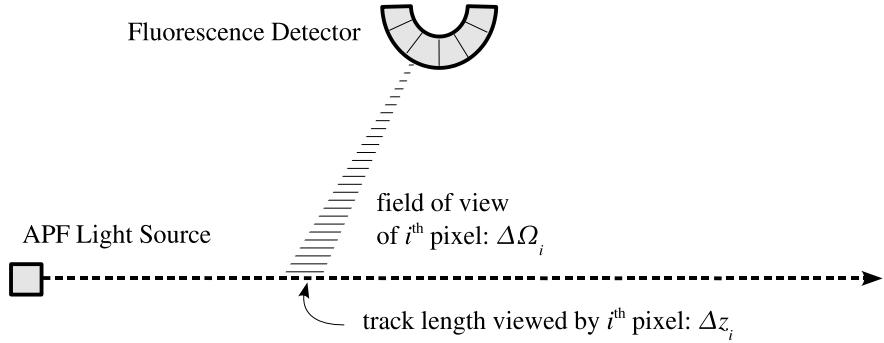}
\caption{\it Schematic of track seen by $i^{th}$ pixel.}
\label{beamgeo}
\vskip1cm
\end{figure}

From the theory of Rayleigh scattering it is known that the Rayleigh phase 
function is 
\begin{equation}
P_{m}(\theta) = \frac{3}{16\,\pi (1+2\,\gamma)}
\left[(1+3\,\gamma)+(1-\gamma)\cos^{2}\theta\right]
\label{eq:rayleigh1}
\end{equation}
where $\gamma$ accounts for the effect of molecular anisotropy on Rayleigh scattering. For
isotropic scattering, $\gamma=0$, this reduces to the familiar 
\begin{equation}
P_{m}(\theta) =  \frac{3}{16\,\pi} (1+\cos^2\theta)~~.
\label{eq:rayleigh2}
\end{equation}
The effect of the anisotropy is small and wavelength-dependent.  Bucholtz~\cite{Bucholtz:1995} 
estimates $\gamma\simeq 0.015$ at 360~nm and concludes that the correction leads to
a $\sim 3\,\%$ systematic increase in the Rayleigh scattering cross section, and a fractional
change $\leq 1.5\,\%$ from the approximate $(1+\cos^{2}\theta)$.  In our analysis, only the
shape of the function is relevant, and we use Eq.\,\ref{eq:rayleigh2} as an approximation
of Eq.\,\ref{eq:rayleigh1}.

The aerosol phase function is often parameterized by the Henyey-Greenstein 
function~\cite{Henyey:1941}:
\begin{equation}
P_{a}(\theta) = \frac{1-g^2}{4\pi} \frac{1}{(1+g^2-2g\mu)^{3/2}}~~,
\label{eq:henyey1}
\end{equation}
where $\mu=\cos\theta$ and $g$ is an asymmetry parameter equal to
the mean cosine of the scattering angle: $g=\langle\cos\theta\rangle$.  The parameter 
$g$ is a measure of how much light is scattered in the forward direction; a greater
$g$ means more light is forward-scattered.  Values for $g$ range from $g=1$ (total 
forward scattering) to $g=-1$ (total backward scattering), with $g=0$ indicating isotropic 
scattering.  

The Henyey-Greenstein function works well for pure forward scattering,
but it cannot describe realistic aerosol conditions, which typically
give rise to non-negligible backscattering.  
Following~\cite{Fishburne:1976,Riewe:1978}, we modify Eq.\,\ref{eq:henyey1}
so that
\begin{equation}
P_{a}(\theta) =
\frac{1-g^2}{4\pi}\left(\frac{1}{(1+g^2-2g\mu)^{3/2}}
+f\frac{3\mu^2-1}{2(1+g^2)^{3/2}}\right)~~.
\label{eq:henyey2}
\end{equation}
The new term in this expression is proportional to the second Legendre
polynomial, and it is introduced to describe the extra backscattering
component.  The value $f$ is a fit parameter used
to tune the relative strength of forward to backward scattering.

The binned APF signal observed in the FD is therefore subjected to a
4-parameter fit:
\begin{equation}
S_i = A \cdot (1+\mu_i^2) + B \cdot (1-g^2) \left(\frac{1}{(1+g^2-2g\mu_i)^{3/2}}
+f\frac{3\mu_i^2-1}{2(1+g^2)^{3/2}}\right)~~, 
\label{eq:fit}
\end{equation}
where $A$, $B$, $g$ and $f$ are the fit parameters.  

In principle, the parameters $A$ and $B$, which describe the relative amount of
Rayleigh and Mie scattering, can be determined from measurements of the extinction 
lengths $\Lambda_m$ and $\Lambda_a$ and assumptions about the particle albedo, 
{\it i.e.} the ratio of light scattered by the aerosol particle in all directions to 
the amount of incoming light.  The albedo is close to one if the particle is mostly 
reflective.  Since local information on the extinction lengths was not available for 
this analysis, we use $A$ and $B$ as additional fit parameters.  We find that the 
distinct shapes of the two phase functions does allow a determination of $A$ and $B$ 
from the data themselves.

\begin{figure}
\centering
\includegraphics[scale=.5]{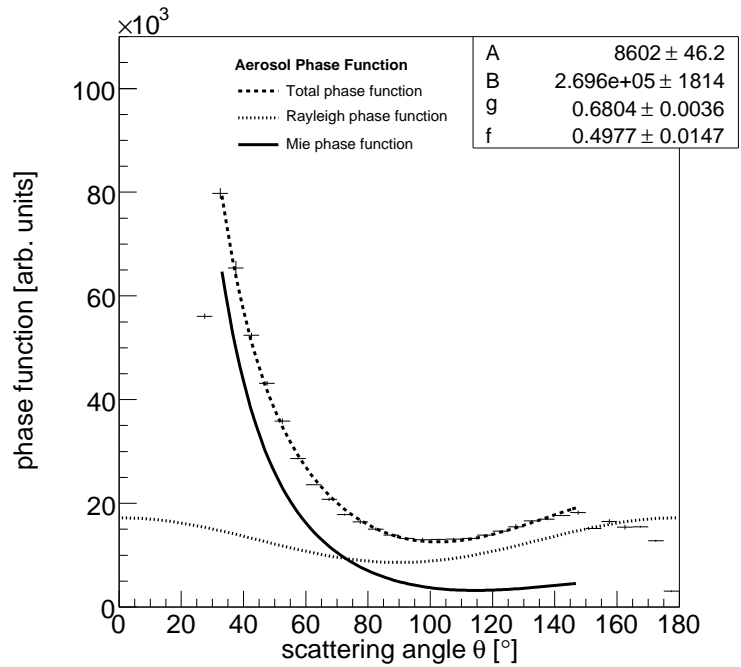}
\includegraphics[scale=.5]{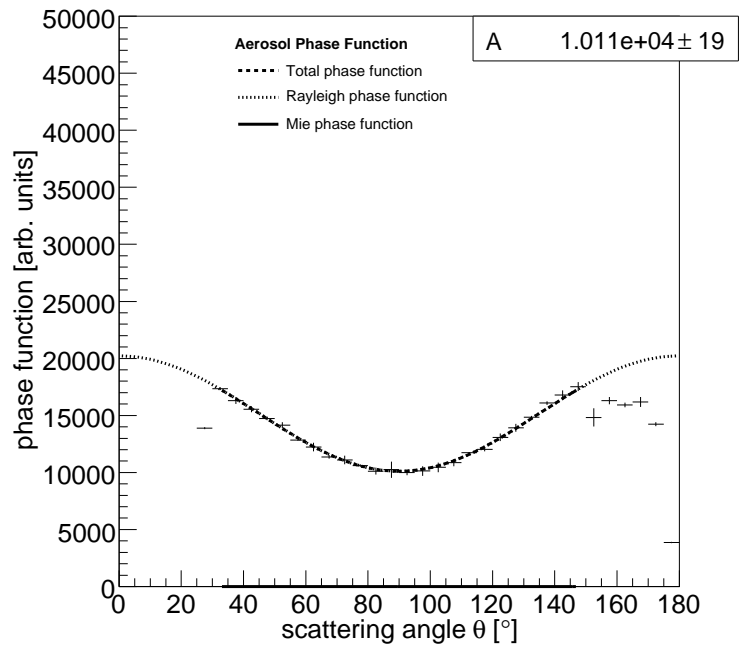} 
\caption{\it Two examples for APF data fits on different days.  In the upper plot
(June 28, 2006, 5:12 am local time) aerosols are visible.  Data are fit to the
function given in Eq.\,\ref{eq:fit}.
The phase function in the lower plot (July 2, 2006, 3:12 am local time)
is consistent with pure Rayleigh scattering.  Data are fit to 
Eq.\,\ref{eq:fit}, with $B=0$, $f=0$, and $g=0$.
Error bars for both plots are the standard deviation of the 5 APF events.}
\label{apf_fits}
\vskip1cm
\end{figure}

At Coihueco, the APF signal is seen in 5 out of the 6 mirrors, so the track is visible
over $\sim 150^{\circ}$ in azimuth.  At the boundary between each mirror there is some 
overlap in the fields of view of pixels.  This overlap produces a double counting 
of signal resulting in the value of bins at boundaries being too large.  These bins are 
simply ignored in the fit.  The values of the other bins and their errors are obtained 
from the mean and standard deviation of the five APF shots in each shot sequence.

On clear nights with few or no aerosols, the fit to Eq.\,\ref{eq:fit} returns unphysical
values for the parameters $B$, $f$, and $g$.  In those cases, we re-fit the data to a 
pure Rayleigh function by setting $B$, $f$, and $g$ equal to zero.  Two examples of
fits, one for a night with aerosol content, and one for a night with pure 
Rayleigh scattering, are shown in Fig.\,\ref{apf_fits}.  The aerosol, molecular, and 
total phase functions are shown.  The aerosol phase function is obtained by subtracting 
the molecular component determined by the fit. 

We fit the data only over a subrange of the available scattering angles,
from $\theta_{min}\simeq 32.5^{\circ}$ to $\theta_{max}\simeq 147.3^{\circ}$.  
As Fig.\,\ref{apf_fits} indicates, the data deviates
from the theoretical prediction for scattering angles below $\theta_{min}$ and above 
$\theta_{max}$.  At smaller and larger angles, several effects corrupt the 
signal and make it unusable for the fit to the phase function.  Due to the local geometry 
at the Coihueco site (see Fig.\,\ref{apf_geo}), the APF shot is not visible for 
$\theta<24^{\circ}$, and below $30^{\circ}$, the signal is incomplete because 
the beam is still partially beneath the detector field of view.
At large scattering angles, the beam is at a rapidly increasing distance to the 
corresponding FD bay, and attenuation of light from the beam to the detector becomes 
important.  As mentioned earlier, because local measurements of the optical depth are 
not yet available, we simply assume $T=1$.  As measurements of $T$ become available, 
the attenuation of light scattered at large angles can be used to correct the data.

In order to apply geometrical corrections when binning the data, the angle at which the 
APF light source shoots ($\gamma$ in Fig.\,\ref{apf_geo}) with respect to the FD and the 
elevation angle of the shot direction needs to be known.  We determined these values from the 
data themselves.  The elevation angle was determined from a reconstruction of APF shots 
with the FD offline reconstruction~\cite{Argiro:2007}, and $\gamma$ was determined from 
the analysis of APF shots on nights where aerosol scattering was negligible.  The data 
from these nights were fit to the Rayleigh component of the phase function, with the 
position of the minimum (nominally at $90^{\circ}$ scattering angle) as a free parameter.  
The fit value of this angle was then used to deduce the direction which the APF light 
source shoots relative to the FD ($\sim 24^{\circ}$ at Coihueco).

\begin{figure}[t]
\centering
\includegraphics[scale=.7]{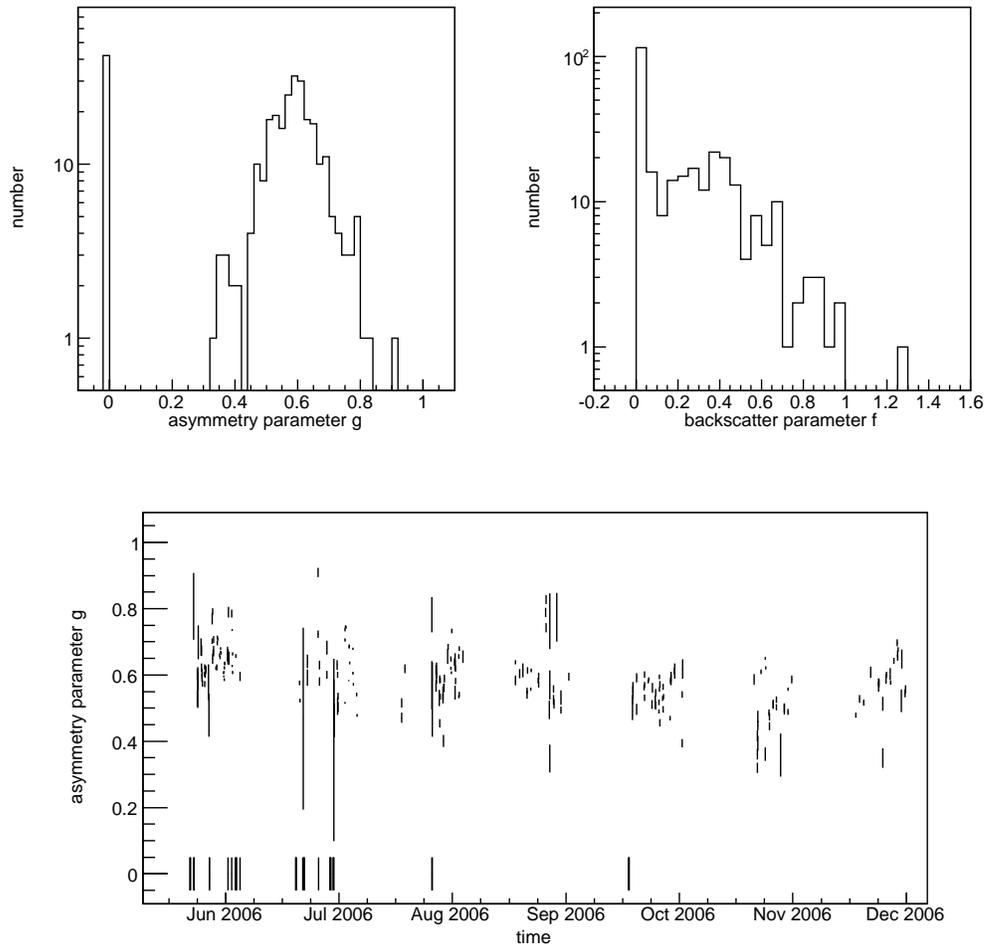} 
\caption{\it {\it Top:} Distribution of the asymmetry parameter $g$ (top left) and 
the backscatter parameter $f$ (top right) for all measurements performed between June
and December 2006.  Values of $g=0$ (and $f=0$) indicate that 
the phase function can be described with pure Rayleigh scattering.  {\it Bottom:} 
Asymmetry parameter $g$ as a function of time.}
\label{gf_param}
\vskip1cm
\end{figure}

\begin{figure}[t]
\centering
\includegraphics[scale=.4]{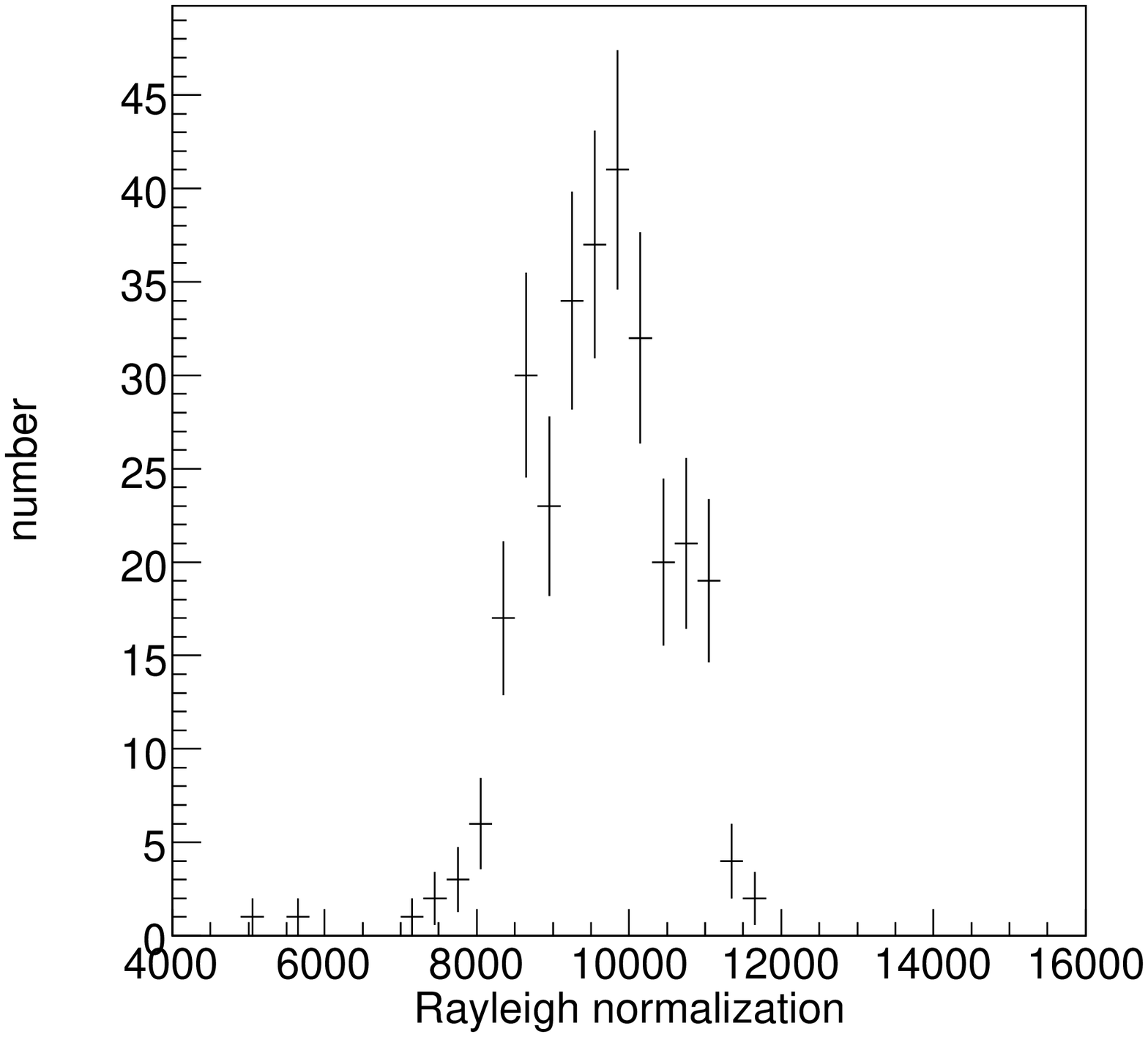} 
\caption{\it Distribution of the Rayleigh normalization parameter $A$
for all measurements performed between June and December 2006.}
\label{A_param}
\vskip1cm
\end{figure}

\section{First Results}

We have applied the analysis described in Section\,\ref{sec:method} to data recorded 
between June and December 2006 at the Coihueco site.  Since the APF light sources
operate during all nights of FD operation, this data set includes all moonless nights,
with the exception of nights with rain or strong winds when the FDs remain closed.
Fig.\,\ref{gf_param} shows the 
distribution of the asymmetry parameter $g$ (left) and the backscatter parameters 
$f$ (right).  For most nights with aerosol contamination, the value of $g$ at the 
experiment site in Malarg\"ue is $\sim 0.6$, with an average of 0.59 and a standard 
deviation of 0.07 for the data period analyzed here.  Values of $g=0$ indicate hours where 
the measured phase function can be described by pure Rayleigh scattering, so the 
aerosol phase function is effectively negligible.  Fig.\,\ref{gf_param} also shows 
the asymmetry parameter as a function of time for the analyzed period.  
With the limited amount of data taken so far, no conclusions concerning seasonal 
variations can be drawn.  The asymmetry parameter appears to be stable during the 
observed time period.  With more data becoming available over the next
few years, we plan to monitor the month-to-month variation in $g$ and
analyze possible correlations with other weather measurements.

One of the main tasks of the APF, in addition to providing the {\it in situ} aerosol 
phase function for every hour of FD data taking, is the identification
of ``clear'' nights with small aerosol contamination.  These nights play an 
important role in the calibration of other atmospheric monitoring devices such as
the Central Laser Facility (CLF)~\cite{Fick:2006}.  On clear nights, the measured 
phase function can be described by pure Rayleigh scattering (measurements where this
is the case appear as $g=0$ in Fig.\,\ref{gf_param}).  

To confirm the reliability of the fit where both the normalization of the Mie and 
the Rayleigh contribution are fit parameters, Fig.\,\ref{A_param} shows the Rayleigh 
normalization factor $A$ for the same data set.  One might expect the molecular
contribution to be rather stable, and in fact this parameter does not change much 
with time.

It is instructive to compare the average asymmetry parameter obtained from the APF
with model expectations and measurements at comparable locations.  Typically, 
measurements are performed at optical wavelengths and cannot be directly compared 
to measurements at UV wavelengths.  However, a compilation at different wavelengths 
from 450~nm to 700~nm~\cite{Fiebig:2006} shows that the wavelength dependence 
of $g$ is small; values at 450~nm are a few percent larger than at 550~nm.  

To first order, $g=0.7$ is often used as a generic value for $g$ in radiative transfer 
models.  A smaller value for $g$ is expected at dry locations.  A parameterization of 
aerosol optical properties by d'Almeida et al.~\cite{dAlmeida:1991} suggests values for $g$ 
between 0.64 and 0.83 at 550~nm depending on aerosol type and season, with higher 
averages for high relative humidity.  

The Pierre Auger Observatory is located east of the Andes in the Pampa Amarilla, 
an arid high plateau at 1420~m a.s.l., so values around 0.6 are within expectations.
For comparison, recent measurements carried out in the 
Southern Great Planes of the US~\cite{Andrews:2006} yield values for $g$ at 550~nm 
of $0.60\pm0.03$ for dry conditions and $0.65\pm0.05$ for ambient conditions.

\begin{figure}[t]
\centering
\includegraphics[width=4in]{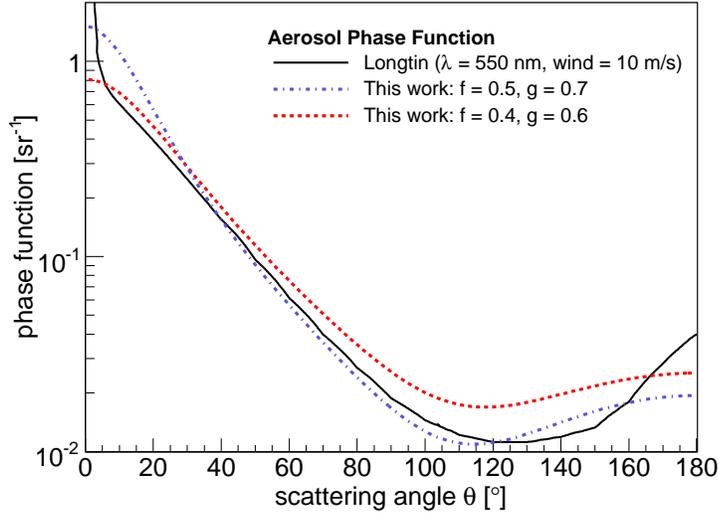}
\caption{\it Comparison of the Longtin aerosol phase function (desert atmosphere
simulated with a wind speed of $10$~m/s) with the default phase
function used in the Auger atmospheric database ($f=0.5$, $g=0.7$)
and the typical phase function measured by the APF ($f=0.4$, $g=0.6$).}
\label{phasefunctions}
\vskip1cm
\end{figure}

The aerosol phase function most commonly used in fluorescence detector data analysis, 
both for the High Resolution Fly's Eye (HiRes) Experiment~\cite{Thomson:2004}, which operated 
in Utah between 1997 and 2006, and the Pierre Auger FD detectors, is the function
obtained from a desert aerosol simulation by Longtin~\cite{Longtin:1988}.
Longtin's desert model is based on Mie scattering theory and assumes that the
desert atmosphere has three major components: carbonaceous particles,
water-soluble particles, and sand. For each aerosol component, the model
assumes a characteristic log normal size distribution and refractive index.
Longtin performed his calculations for several wavelengths and wind speeds;
those made at 550~nm with a wind 
speed of 10~m/s most closely match the 300~nm to 400~nm nitrogen 
fluorescence band observed by the FDs and have therefore been traditionally 
used in air fluorescence data analysis.

Fig.\,\ref{phasefunctions} compares the Longtin aerosol phase function at 550~nm 
to the modified Henyey-Greenstein function of Eq.\,\ref{eq:henyey2} with two sets 
of $f$ and $g$: $f=0.5$ and $g=0.7$, the default values used by the Auger
atmospheric database; and $f=0.4$ and $g=0.6$, the values determined in this
study to be more typical of the detector location.  The comparison shows that,
on average, the difference between the Longtin function and the measured phase 
function is small for those scattering angles relevant in fluorescence measurements
--- $\sim 30^\circ$ to $150^\circ$.  Only at the largest scattering angles above 
$160^\circ$ do the phase functions differ notably.  This region is outside the 
current range of validity of our measurement.

\begin{figure}[t]
\centering
\includegraphics[width=4.5in]{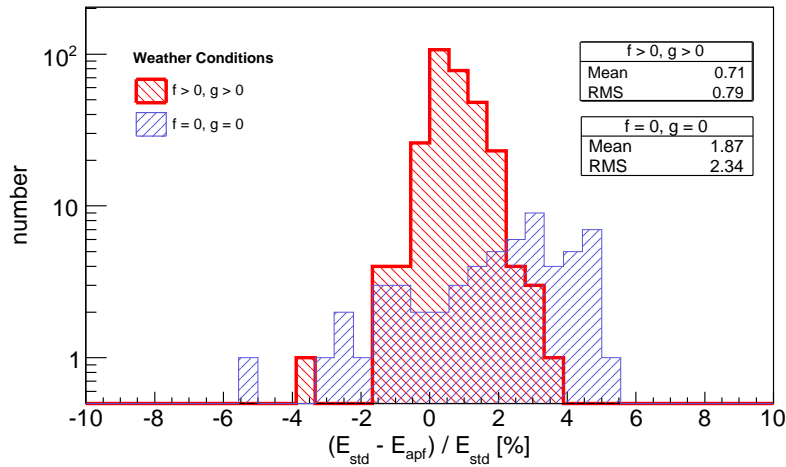}
\caption{\it Differences in the energies of golden hybrid events reconstructed with
default phase function values $(E_\text{std})$ and those reconstructed
using phase function fit parameters determined from APF measurements
$(E_\text{apf})$.  The red (bold) histogram represents data taken during
nights with measurable aerosols; the blue (light) histogram depicts events
observed on purely molecular nights.}
\label{energyspread}
\vskip1cm
\end{figure}

Our primary interest in aerosol scattering is its effect on the air shower reconstruction,
most notably the determination of the shower energy.  However, it is not straightforward to
estimate the extent to which the use of measured rather than averaged values of $f$ and $g$
changes the energy reconstruction, as this depends strongly on other atmospheric parameters,
for example the aerosol optical depth.  Rather than singling out the phase function
measurement, we need to study the effect of the combined measurement of all atmospheric 
parameters, a task which is beyond the scope of this paper.

We can, however, get an estimate of the relevance of the phase function measurement     
by studying its effect on the energies of events that are of particular importance 
for the energy calibration of the detector, the ``golden hybrid events.''  These are 
events observed by one or more fluorescence detectors and three or more surface array 
tanks.  For ``golden hybrid events'' observed by the Coihueco FD site between June and
December 2006, we performed the reconstruction twice: first, using the default
parameters $f=0.5$ and $g=0.7$ to estimate aerosol scattering; and second,
using the fit parameters determined from APF measurements.  In both cases
atmospheric extinction was simulated using an average aerosol profile model
representative of the Malarg\"ue site~\cite{BenZvi:2007a,Prouza:2007}.  

Fig.\,\ref{energyspread} depicts the relative differences in energies
caused by reconstructing showers with the default phase function and the
measured phase function.  The red (bold) histogram represents data taken during
nights with aerosol contamination ($f>0$, $g>0$) while the blue (light) histogram
represents data taken during nights where according to the APF analysis
scattering is purely molecular.  The correction is typically of order one percent.
However, on those nights when aerosol loading is extremely low,
so that atmospheric scattering may be characterized as purely molecular, the
use of the default scattering parameters causes
larger errors in the shower reconstruction.  Under such conditions, the
total phase function lacks the strong forward-scattering component typical of
aerosols.  During these periods, incorrectly accounting for aerosol scattering 
starts to impact the energy calibration of the detector.  A correct determination 
of the phase function on a regular basis is
therefore an important part of the atmospheric monitoring efforts at the site.

\section{Conclusions and Outlook}

As part of the atmospheric monitoring program at the Pierre Auger Observatory, 
the aerosol phase function at 350~nm is routinely measured
at two of the four FD sites.  A first analysis of data taken from June to December 2006
shows that values of $g=\langle\cos\theta \rangle\simeq 0.6$ for the mean cosine of the
scattering angle $\theta$ are typical for aerosols at the site of the experiment.  
Over the next several years, the APF light sources will
produce a data set of unprecedented size of the scattering properties of aerosols.
This data set will enable us to carefully study any seasonal change in the 
aerosol content.  The APF light sources and the other atmospheric monitoring instruments
at the Auger site will accumulate one of the largest sets of continuous measurements in the 
300~nm to 400~nm range ever recorded for a single location.

The APF light sources are currently operating at a wavelength of 350~nm only.  In the near 
future, we will add regular measurements at 330~nm and 390~nm to study the dependence 
of the phase function on the wavelength of the scattered light.

\ack
We are grateful to the following agencies and organizations for financial support:
The APF light sources were built by a grant from the 
Department of Energy (DOE) Office of Science (USA) (DE-FG03-92ER40732).
Parts of the APF analysis were performed during the 2006 REU (Research Experience for
Undergraduates) program at Columbia University's Nevis Laboratories which is supported
by the National Science Foundation (USA) under contract number NSF-PHY-0452277.

\end{document}